\documentclass[journal=jacsat,manuscript=communication,layout=twocolumn]{achemso}

\usepackage[version=3]{mhchem} 
\usepackage[T1]{fontenc}       

\author{L. Banszerus}
\affiliation{JARA-FIT and 2nd Institute of Physics, RWTH Aachen University, 52074 Aachen, Germany}

\author{H. Janssen}
\affiliation{JARA-FIT and 2nd Institute of Physics, RWTH Aachen University, 52074 Aachen, Germany}

\author{M. Otto}
\affiliation{Advanced Microelectronic Center Aachen (AMICA), AMO GmbH, Otto-Blumenthal-Stra\ss e 25, 52074 Aachen, Germany}

\author{A. Epping}
\affiliation{JARA-FIT and 2nd Institute of Physics, RWTH Aachen University, 52074 Aachen, Germany}
\alsoaffiliation{Peter Gr\"unberg Institute (PGI-9), Forschungszentrum J\"ulich, 52425 J\"ulich, Germany}

\author{T. Taniguchi}
\affiliation{National Institute for Materials Science, 1-1 Namiki, Tsukuba 305-0044, Japan}

\author{K. Watanabe}
\affiliation{National Institute for Materials Science, 1-1 Namiki, Tsukuba 305-0044, Japan}

\author{B.~Beschoten}
\affiliation{JARA-FIT and 2nd Institute of Physics, RWTH Aachen University, 52074 Aachen, Germany}

\author{D. Neumaier}
\affiliation{Advanced Microelectronic Center Aachen (AMICA), AMO GmbH, Otto-Blumenthal-Stra\ss e 25, 52074 Aachen, Germany}

\author{C. Stampfer}
\affiliation{JARA-FIT and 2nd Institute of Physics, RWTH Aachen University, 52074 Aachen, Germany}
\alsoaffiliation{Peter Gr\"unberg Institute (PGI-9), Forschungszentrum J\"ulich, 52425 J\"ulich, Germany}
\email{stampfer@physik.rwth-aachen.de}

	\title
  {Identifying suitable substrates for high-quality graphene-based heterostructures}


\begin{document}


\newpage
\begin{abstract}
We report on a scanning confocal Raman spectroscopy study investigating the strain-uniformity and the overall strain and doping of high-quality chemical vapour deposited (CVD) graphene-based heterostuctures on a large number of different substrate materials, including hexagonal boron nitride (hBN), transition metal dichalcogenides, silicon, different oxides and nitrides, as well as polymers. By applying a hBN-assisted, contamination free, dry transfer process for CVD graphene, high-quality heterostructures with low doping densities and low strain variations are assembled. The Raman spectra of these pristine heterostructures are sensitive to substrate-induced doping and strain variations and are thus used to probe the suitability of the substrate material for potential high-quality graphene devices. We find that the flatness of the substrate material is a key figure for gaining, or preserving high-quality graphene.

\end{abstract}

{\textbf{Introduction}}\\
For over a decade, graphene has been in the spotlight of solid state research. Its high charge carrier mobilities\cite{Ban16,Wan13,Ban15,Bol08} and long spin diffusion lengths\cite{Dro16,Ing15}, as well as its optical\cite{Bon10} and mechanical properties\cite{Lee07} promise a wide range of applications ranging from spintronics\cite{Roc15} to high frequency electronics\cite{Neu15}, ultra-sensitive sensors\cite{Dau16, Wan16} and flexible optoelectronics \cite{Fer15}. In order to advance prototype devices to true applications, large effort has been put into growth\cite{Li09,Che13,Bae10,Li11,Lee14} and contamination-free transfer\cite{Suk11,Pet12,Ban15,Ban16} of high quality graphene based on chemical vapour deposition.
However, as graphene and other two-dimensional (2d) materials consist only of surface atoms, the choice of substrate material has a large influence on their structural and electronic properties\cite{Gan11,Kre14,Dea10,Wan13,Cuo14, Pir11}.
In this work, we investigate strain, doping and the strain uniformity of high quality CVD graphene/hBN heterostructures placed on different substrate materials.
Here, we follow a recently reported, contamination free, dry transfer process, where exfoliated hBN is used to pick up CVD graphene directly from the growth substrate. The obtained stack is subsequently placed on different target substrates\cite{Ban15}.
This fabrication process yields high quality heterostructures with little intrinsic overall doping and low nanometre-scale strain variations.
As the graphene crystal is covered, i.e. protected by hBN on the top side, modifications in doping and strain are purely
due to the substrate at the bottom side of graphene, making our approach suitable for benchmarking the substrate suitability.   \\
\begin{figure}[t]\centering
\includegraphics[draft=false,keepaspectratio=true,clip,%
                   width=\linewidth]%
                   {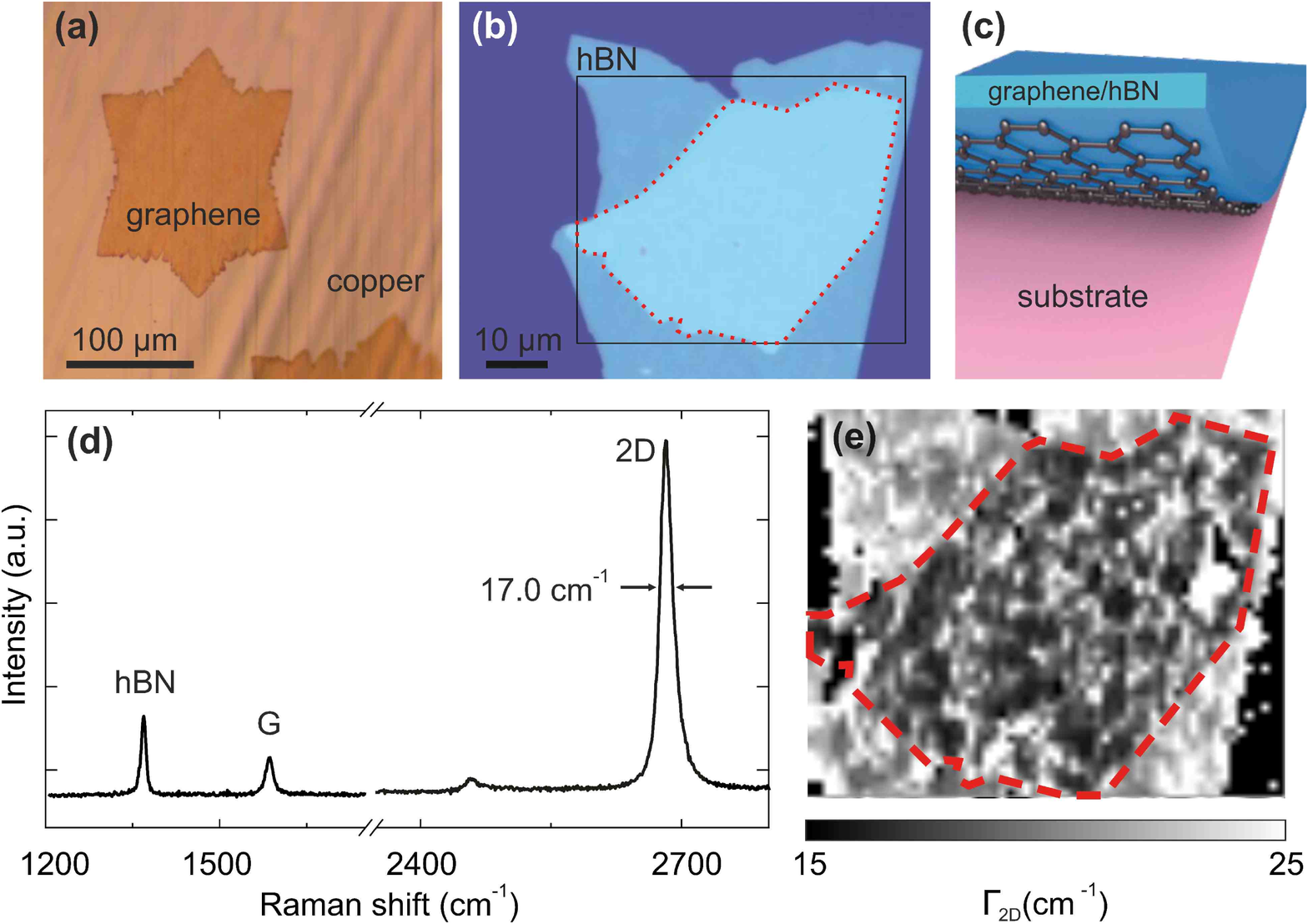}
\caption[fig01]{ (a) Optical image of a graphene crystal on copper foil a few days after growth. The orange colour is due to the oxidised interface between graphene and the subjacent copper. (b) Optical microscope image of a hBN/CVD-graphene/hBN heterostructure. The black rectangle marks the area that has been mapped using confocal Raman microscopy. The red dashed line marks the sandwiched area. (c) Schematic of a hBN/graphene/substrate structure. (d)  Typical Raman spectrum of CVD graphene encapsulated between two flakes of hBN. (e) Raman map of the line width of the Raman 2D-peak, $\Gamma_{\mathrm{2D}}$ of the area marked in panel (b).}
\label{fig01}
\end{figure}
\begin{figure*}[t]\centering
  \includegraphics[draft=false,keepaspectratio=true,clip,%
                     width=0.85\linewidth]%
                     {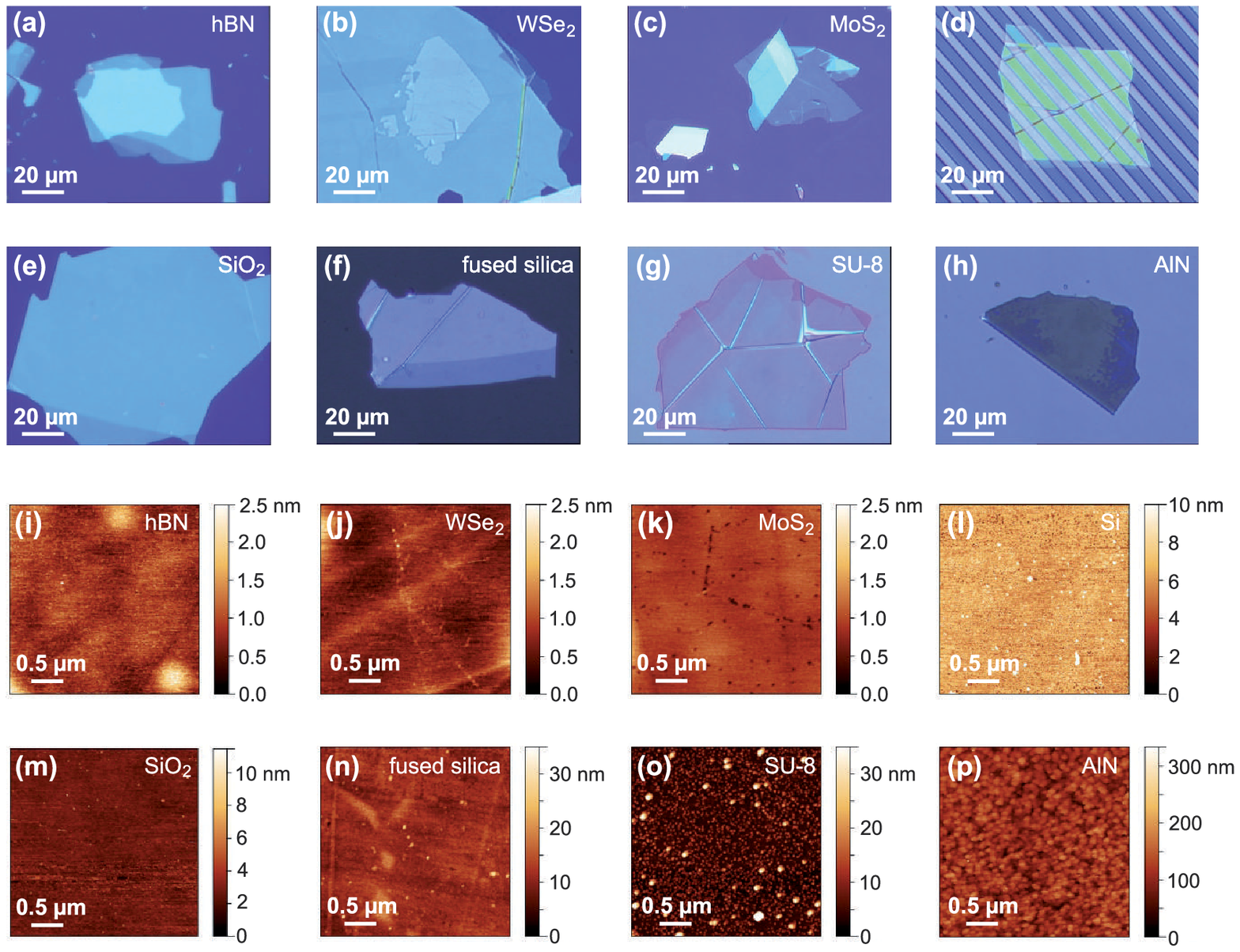}
  \caption[fig02]{Optical microscope images of the hBN/graphene/substrate heterostructures based on different substrates including (a) hBN, (b) WSe$_2$, (c) MoS$_2$, (d) graphene hBN suspended over trenches, (e) Si$^{++}$/SiO$_2$, (f) fused Silica (g) SU-8, and (h) aluminum nitride (AlN). Atomic force microscopy images of the substrate materials including (i) hBN, (j) WSe$_2$, (k) MoS$_2$, (l) Silicon, (m) Si$^{++}$/SiO$_2$, (n) fused Silica (o) SU-8, and (p) aluminium nitride (AlN). }
  \label{fig02}
  \end{figure*}\\

{\textbf{Sample fabrication}}\\
We grow individual graphene crystals with a typical diameter of a few hundred micrometers using a low pressure chemical vapour deposition process at a growth temperature of 1035$^\circ$C on the inside of enclosures folded from commercial copper foil (AlfaAesar 46365). Growth is carried out in a hydrogen/methane atmosphere (45~sccm and 5~sccm respectively) at a total pressure of 0.1~mbar for two hours.\cite{Che13}. Similar to previous works, we let the interface between graphene and copper oxidise under ambient conditions for a few days prior to transfer, in order to weaken the adhesion of the graphene to the copper foil\cite{Ban15,Ban16}. Figure 1(a) shows an optical microscope image of an individual graphene crystal a few days after growth. The bright orange colour is caused by the copper oxide layer on the copper-to-graphene interface. In order to transfer the graphene from the copper foil onto a different substrate, we employ a contamination-free van-der-Waals dry-transfer process. For the transfer, a polymer stamp consisting of a polydimethylsiloxane (PDMS) cushion is covered with a  polyvinylalcohol (PVA)/ polymethylmethacrylate (PMMA) stack with a flake of exfoliated hBN on top. The hBN flake is aligned with the graphene and brought into contact at a temperature of 125$^\circ$C and separated again from the copper foil. Subsequently, the graphene/hBN stack is placed on different target substrates and the polymers are dissolved in deionized-water, acetone and isopropanol.\\
 \begin{figure*}[htb!]\centering
 \includegraphics[draft=false,keepaspectratio=true,clip,%
                    width=0.85\linewidth]%
                    {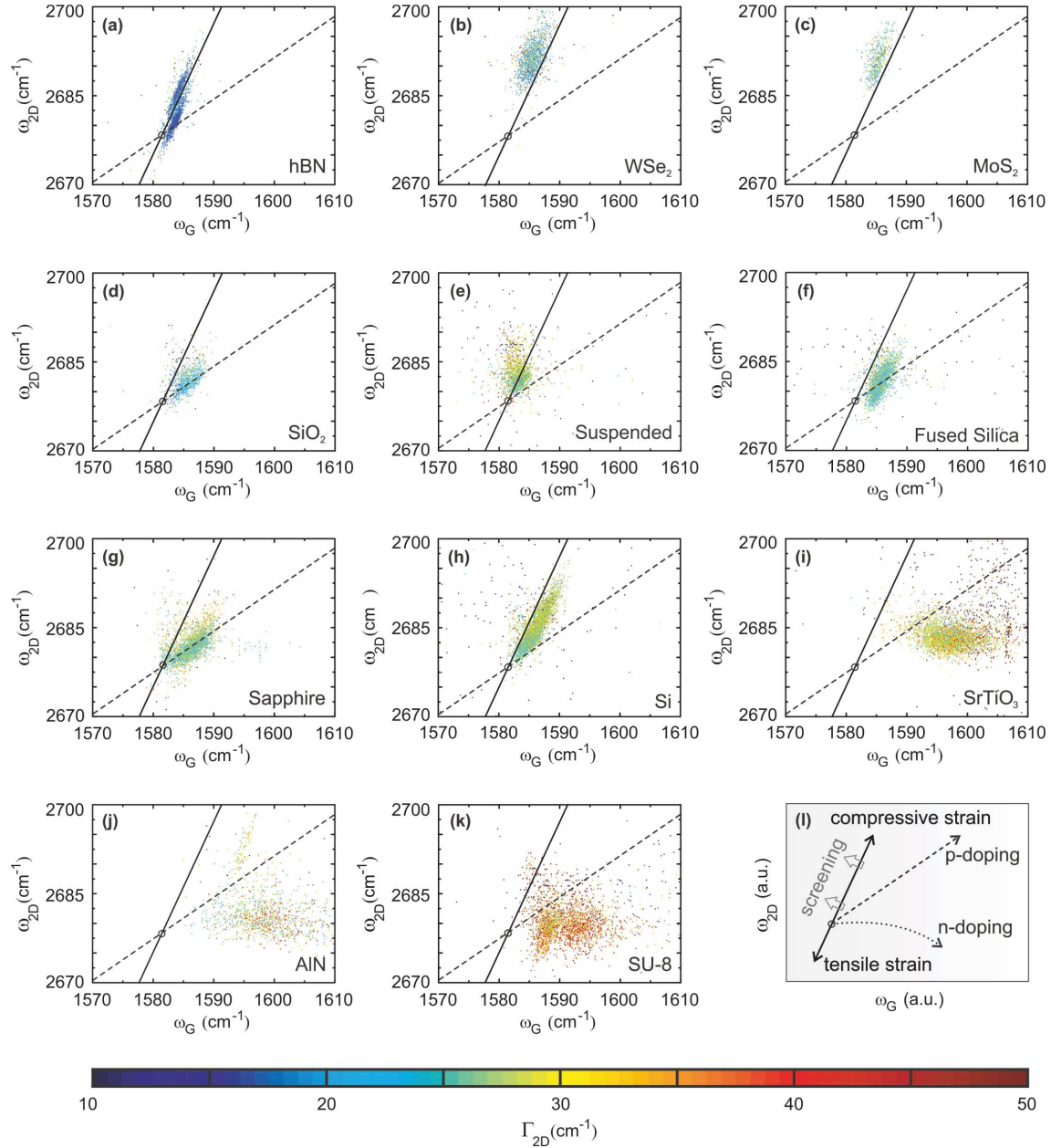}
 \caption[fig03]{(a)-(k) Scatter plots of $\omega_\mathrm{2D}$ vs. $\omega_\mathrm{G}$ extracted from the spatially resolved Raman maps recorded on different substrates. Strain and doping of the graphene can be extracted using a kind of vector decomposition model\cite{Lee12}. The black dashed lines correspond to the doping axes and the black solid lines correspond to the strain axes. The black circles represent the point of pristine unstrained graphene. (l) Schematic representation of the influence of strain, p- and n-doping, as well as dielectric screening on  $\omega_\mathrm{2D}$ and $\omega_\mathrm{G}$. }
 \label{fig03}
 \end{figure*}
	
  {\textbf{Results and discussion}}\\
 Figure 1(b) shows an optical microscope image of a graphene/hBN stack placed on another flake of hBN. In figure~1(c), we illustrate a schematic of such a hBN/graphene stack on an arbitrary substrate. In previous works, we reported on very little overall doping concentrations ($n < 3\times 10^{11}$~cm$^{-2}$) and charge carrier mobilities in the range of millions of cm$^2$/(Vs) for similar hBN/graphene/hBN structures\cite{Ban16}, indicating that the transfer process avoids degradations of the graphene quality\cite{Ban15,Ban16}. In order to investigate the quality of the resulting substrate/graphene/hBN stacks, we use scanning confocal Raman microscopy, which is a fast and non-invasive method for probing a number of material properties of graphene including strain, doping, defects and nanometre-scale strain variations\cite{Fer06,Fer13,Gra07,Lee12,Mal09,Neu14,For13}. Confocal Raman microscopy was carried out using a WiTec 300 system using a laser with 532~nm wavelength and a typical laser power of 1~mW. Regions with cracks and folds in the hBN are disregarded in the analysis. Figure 1(d) shows a typical Raman spectrum of graphene sandwiched between two flakes of hBN. The Raman peak originating from the hBN is located around $\omega_{\mathrm{hBN}}$~=~1365~cm$^{-1}$. The absence of the Raman D-peak at 1345~cm$^{-1}$ indicates a very low defect density in the transferred CVD graphene. The Raman G-peak located around $\omega_{\mathrm{G}}$~=~1582~cm$^{-1}$ shows a full-width-at-half-maximum (FWHM), $\Gamma_{\mathrm{G}}$, of around 14.5~cm$^{-1}$, indicating very little overall doping of the graphene-based heterostructure\cite{Cas07}. The Raman 2D-peak is located around $\omega_{\mathrm{2D}}$~=~2678~cm$^{-1}$. The FWHM of the 2D-peak, $\Gamma_{\mathrm{2D}}$, is 17~cm$^{-1}$ indicating a high sample quality\cite{Ban15}.
Recently, it has been shown that the value of $\Gamma_{\mathrm{2D}}$ is related to the amount of nanometre-scale strain variations in the graphene lattice within the spot size of the laser\cite{Neu14}.
This is particularly crucial as there are strong indications that nanometre-scale strain variations are a limiting factor for the charge carrier mobility in high-quality graphene\cite{Cuo14}.
This stresses the importance of using the line width of the Raman 2D-peak in order to characterise potential substrates for graphene. Figure 1(e) shows a Raman map of $\Gamma_{\mathrm{2D}}$ recorded in the area marked by the black box in figure 1(b).
Here, $\Gamma_{\mathrm{2D}}$ is homogeneously distributed around 18~cm$^{-1}$ over the entire sandwiched region of the sample [see red dashed line in figures 1(b) and~1(e)], indicating a homogeneous sample quality with little strain variations over the entire sandwiched area. Outside the sandwiched area where the hBN/graphene stack lays on SiO$_2$,  $\Gamma_{\mathrm{2D}}$ shows elevated values of around 23~cm$^{-1}$, revealing substrate-induced strain variations in the graphene.

The data extracted from the collected Raman spectra on the CVD graphene/hBN sandwich area are very similar to those of high quality heterostructures obtained with exfoliated graphene\cite{Neu14,Dea10,Wan13}, showing that neither the CVD growth nor our dry transfer method are limiting factors for the 'structural' quality (i.e. overall, 'microscopic' strain and nanometre-scale strain variations) and doping of the graphene-based heterostructures.

In order to investigate and identify suitable other (potentially more scalable) substrate materials, we fabricate stacks of graphene/hBN heterostructures and place them on a number of different substrates. Figures 2(a)-2(h) show optical images of graphene/hBN stacks on such different substrate materials. The substrates investigated in this study are hBN, transition metal dichalcogenides (TMDCs) such as WSe$_2$ and MoS$_2$, Si$^{++}$/SiO$_2$, sapphire, SrTiO$_3$, graphene/hBN suspended over trenches, silicon, hydrogen-terminated fused silica, aluminium nitride (AlN) and  cross-linked SU-8 (negative epoxy resist from MicroChem). These substrates cover the material classes of oxides, nitrides, silicon, polymers and 2d-materials. With this selected set of substrate materials, we investigate in particular technologically relevant substrate materials as well as upcoming 2d materials, which have already been shown to be good and potentially scalable substrates for graphene\cite{Kre14}.
In order to understand the influence of the surface roughness of the substrate on the structural properties of graphene, the substrate roughness is measured using atomic force microscopy (AFM). Figures~2(i)-2(p) show AFM images of the different substrates used. The AFM maps are used to extract the root mean square (RMS) values of the surface roughness for the individual substrate materials. The 2d-materials hBN, WSe$_2$ and MoS$_2$ show the smallest roughness with RMS values less than 0.2~nm. Silicon, sapphire and Si$^{++}$/SiO$_2$ show RMS values between 0.5~nm and 1~nm. Hydrogen-terminated fused silica, SrTiO$_3$ and SU-8 have a roughness between 2~nm and 4~nm and the AlN substrate has the largest roughness with an RMS value of 26~nm. Kretinin and co-workers\cite{Kre14} report RMS values of 0.1~nm for hBN and TMDCs and around 1~nm for the oxides they investigated, which compares well with the values we observe.
Independently, it is important to note that the measured RMS roughness values are not intrinsic substrate properties, but are generally influenced by the used deposition method and subsequent substrate treatments.

\begin{figure}[t]\centering
\includegraphics[draft=false,keepaspectratio=true,clip,%
                   width=0.75\linewidth]%
                   {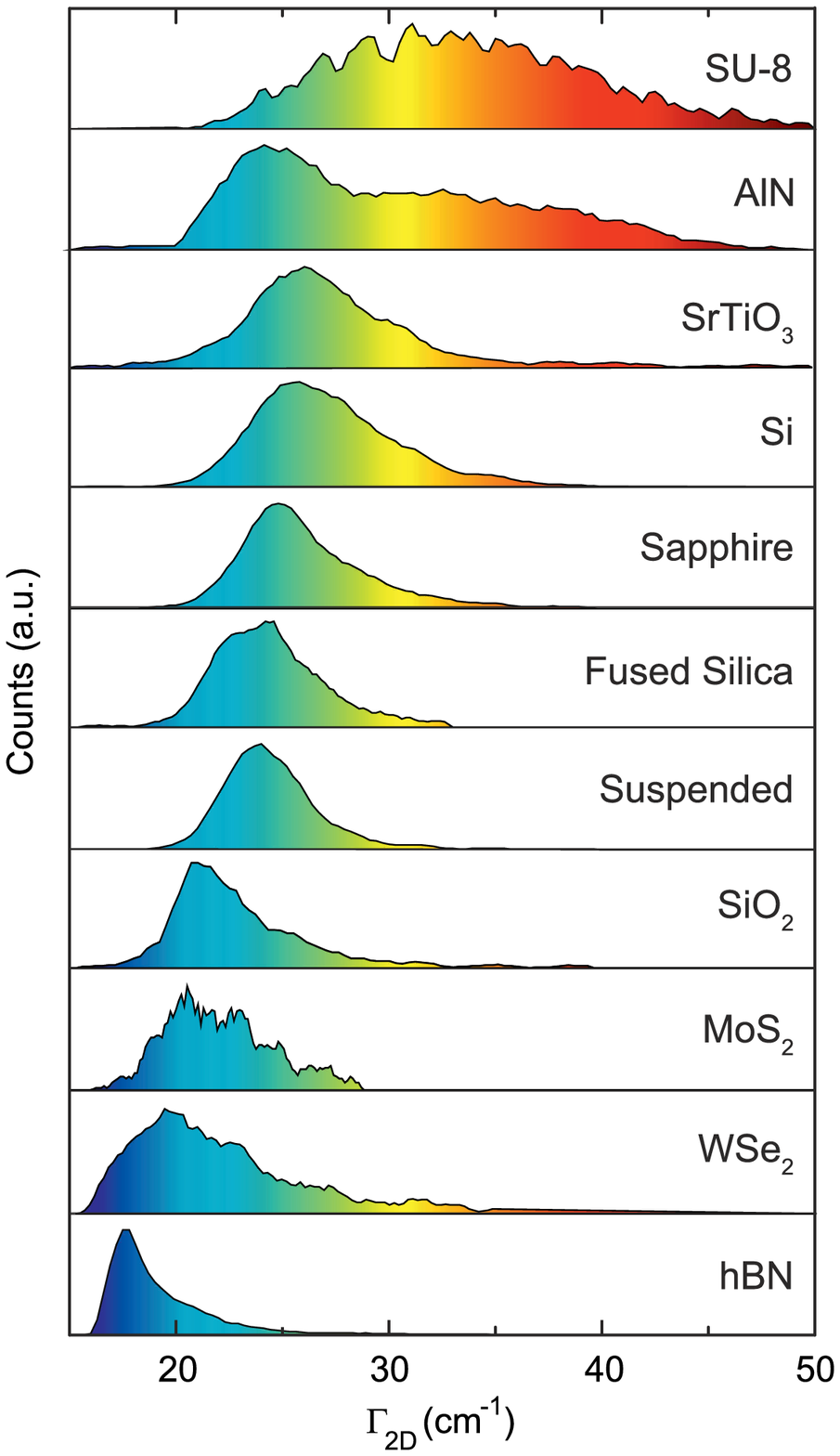}
\caption[fig04]{ Histograms of $\Gamma_{\mathrm{2D}}$ values extracted from the Raman maps taken on the graphene/hBN heterostructures placed on the different substrates investigated. The colour code is adapted from figure 3. }
\label{fig04}
\end{figure}

In figure~3, we summarize all the information about substrate-induced strain, doping and nanometre-scale strain variations obtained from the Raman maps of graphene/hBN on each substrate material by showing scatter plots of $\omega_{\mathrm{2D}}$ and $\omega_{\mathrm{G}}$. The amount of nanometre-scale strain variations is linked to $\Gamma_\mathrm{2D}$, which is colour coded in the data points in each panel. In particular, we observe that graphene on hBN, WSe$_2$, MoS$_2$, and SiO$_2$ shows low nanometre-scale strain variations as seen by the small values of $\Gamma_\mathrm{2D}$ (blue colour of data points), whereas substrates like SrTiO$_3$, AlN and SU-8 show very large values of $\Gamma_\mathrm{2D}$. Besides nanometre-scale strain variations, the overall doping (which also is connected to doping variations) strongly influences the transport properties of graphene. In order to compare the amount of substrate-induced strain and doping in graphene for the different substrates, we follow the method reported by Lee et al.\cite{Lee12} and employ a 'vector' decomposition model on the position of the G-peak and the 2D-peak. A schematic of the influence of strain, doping and dielectric screening of the substrate is depicted in figure~3(l). Strain shifts the 2D-peak and G-peak positions with a relative slope of 2.2, which is related to the ratio of the Gr\"uneisen parameters for the 2D-peak and the G-peak phonons\cite{Neu14,Lee12} [see black solid line in figure 3(l) and all other panels of figure 3]. Hole doping shifts $\omega_\mathrm{2D}$ and $\omega_\mathrm{G}$ with a relative slope of 0.7 [black dashed line in figure 3(l) and all other panels of figure 3] and n-doping results in a non-linear decrease of $\omega_\mathrm{2D}$\cite{Lee12,Fro15} [see black dotted line in figure~3(l)]. The point of zero strain and zero doping is marked by the black circle. We applied a method by Berciaud et al.\cite{Ber13} and Lee et al.\cite{Lee12} in order to determine this point by measuring Raman spectra of pristine suspended graphene. The graphene was in this case directly exfoliated onto a Si/SiO$_2$ chip with holes etched into the substrate (sample not shown). The G-peak and the 2D-peak of this pristine graphene sample are located at $\omega_{\mathrm{G}}=1581.6~\pm 1$~cm$^{-1}$ and $\omega_{\mathrm{2D}}=2678.6~\pm 1$~cm$^{-1}$, respectively. In addition to strain and doping, dielectric screening of the substrate material has been shown to shift the peak positions of the G-peak and the 2D-peak as represented by the arrows in figure~3(l)\cite{For13}. \\

Now we have everything to read the information encoded in figures 3(a)-(k). We first compare how far the data points are shifted from the strain axis (black solid line). In particular offsets to the right indicate a minor electronic quality as they are caused by significant doping and doping variations leading to an increased Coulomb scattering. In contrast, small offsets to the left of the black solid line are caused by substrate-induced screening effects, not affecting the (electronic) quality of graphene. All this allows to interpret our data, leading e.g. to the conclusion that WSe$_2$ and MoS$_2$ (only left shifts) are better substrates than SiO$_2$ or sapphire (doping present), even though both substrate classes show almost the same amount of nanometre-scale strain variations. Furthermore, it becomes obvious that SrTiO$_3$, AlN and SU-8 not only introduce significant nanometre-scale strain variations, but also lead to substantial charge carrier doping in graphene, making these materials not very suitable for high-quality graphene devices. Notably, the graphene suspended over trenches shows a slightly broadened distribution of $\omega_{\mathrm{2D}}$ and $\omega_{\mathrm{G}}$ and elevated values for $\Gamma_{\mathrm{2D}}$, which most likely results from an transfer related contaminations, degrading the graphene quality through the open trenches in the substrate.\\

Figure~4 shows the individual histograms of the $\Gamma_{\mathrm{2D}}$ values extracted from Raman maps of graphene/hBN on all substrates investigated. The histograms for hBN and WSe$_2$ and MoS$_2$ show very low values of  $\Gamma_{\mathrm{2D}}$, indicating a very low amount of nanometre-scale strain variations, which makes these 2d materials suitable substrates for graphene. Interestingly, these materials exhibit a sharp lower cut-off at around 16~cm$^{-1}$ suggesting this to be the intrinsic, non-broadened line width of the Raman 2D mode for our laser wavelength. In contrast, the oxidic materials such as SiO$_2$, sapphire and SrTiO$_3$ show elevated values of $\Gamma_{\mathrm{2D}}$. Both, AlN and SU-8 show extremely high values of $\Gamma_{\mathrm{2D}}$ of up to 50~cm$^{-1}$ in combination with broad distributions suggesting not only a very high amount of nanometre-scale strain variations, but also non-homogeneous sample properties. \\

\begin{figure}[tb!]\flushright
\includegraphics[draft=false,keepaspectratio=true,clip,%
                   width=\linewidth]%
                   {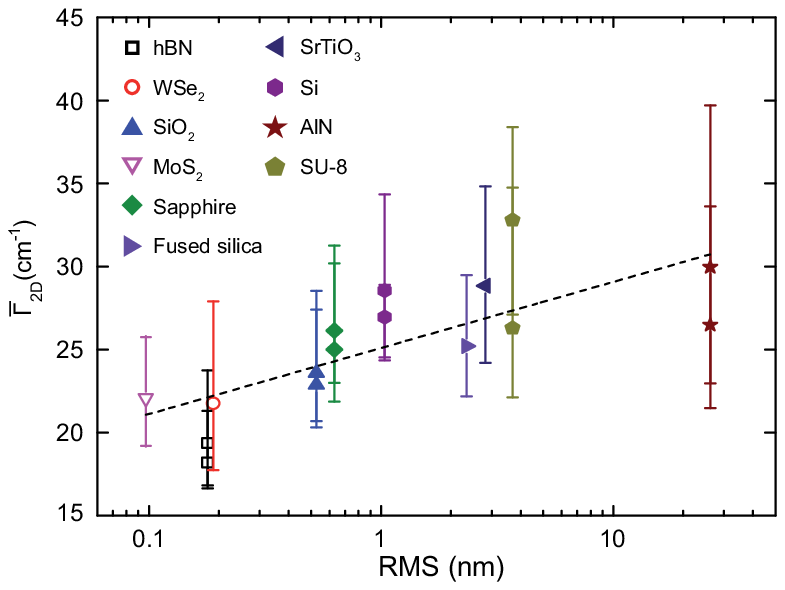}
\caption[fig05]{$\overline{\Gamma}_\mathrm{2D}$ versus substrate roughness. The open data points mark the 2d-layered materials that show self cleaning properties. The black line is a guide to the eye, indicating an increasing $\Gamma_{\mathrm{2D}}$ for increasing surface roughness. }
\label{fig05}
\end{figure}
In figure~5, we plot the sample averaged $\overline{\Gamma}_\mathrm{2D}$ versus the RMS values of the surface roughness of the substrate material to explore how the presence of nanometre-scale strain variations is related to the surface roughness. The error bars plotted in figure 5 indicate the 20th and the 80th percentile of the corresponding distributions of $\Gamma_\mathrm{2D}$. The hollow data points belong to the 2d materials, which exhibit a self cleaning effect\cite{Kre14}. The black dashed line is a guide to the eye, indicating an increase of the sample averaged $\Gamma_\mathrm{2D}$ value as function of an increasing RMS value of the surface roughness. This observation suggest that very flat substrate materials induce very little nanometre-scale strain variations in graphene and - as long as they do not induce large doping/doping variations - are well suited as substrate for high mobility graphene\cite{Cuo14}. A similar relation between $\overline{\Gamma}_\mathrm{2D}$ and the surface roughness has been observed for the case of graphene on metallic substrates\cite{Zha14}. However, in contrast to Zhao et al.\cite{Zha14}, we observe a smaller variation of $\overline{\Gamma}_\mathrm{2D}$ as function of substrate roughness, which may result from the fact that the graphene/hBN heterostack is stiffer than a single layer of graphene and is thus less susceptible to strain than a single layer of graphene on a rough substrate.

Table 1 summarizes our findings including typical RMS values of the surface roughness, the sample averaged values of the FWHM of the 2D peak, $\overline{\Gamma}_{\mathrm{2D}}$ and the mean values for the position of the G-peak $\overline{\omega}_{\mathrm{G}}$ and of the 2D-peak $\overline{\omega}_{\mathrm{2D}}$ for all substrate materials investigated.

\begin{table}[t]
\centering
\begin{tabular}{|p{1.4cm}||p{1cm}|p{1.2cm}|p{1.2cm}|p{1.2cm}|}
\hline  & RMS (nm) & $\overline{\Gamma}_\mathrm{2D}$ (cm$^{-1}$)& $\overline{\omega}_\mathrm{G}$ (cm$^{-1}$) & $\overline{\omega}_\mathrm{2D}$ (cm$^{-1}$) \\
\hline
\hline hBN & 0.17 & 18.2 & 1583.8 & 2682.6 \\
\hline WSe$_2$ & 0.19 & 21.8 & 1587.2 & 2690.9 \\
\hline MoS$_2$ & 0.09 & 22.0 & 1585.1 & 2690.7 \\
\hline SiO$_2$ & 0.53 & 22.8 & 1585.5 & 2682.5 \\
\hline susp. & --- & 24.5 & 1584.2 & 2682.3 \\
\hline fus. silica & 2.3 & 25.2 & 1585.6 & 2681.7 \\
\hline sapphire & 0.63 & 26.1 & 1587.1 & 2682.7 \\
\hline silicon & 1.0 & 26.9 & 1583.2 & 2685.9 \\
\hline SrTiO$_3$ & 2.8 & 28.8 & 1596.9 & 2683.4 \\
\hline AlN & 26.2 & 30.0 & 1600.3 & 2681.7 \\
\hline SU-8 & 3.7 & 33.6 & 1591.2 & 2681.0 \\
\hline
\end{tabular}
\caption{Representation of the surface roughness (RMS), the sample averaged values of the FWHM of the 2D peak, $\overline{\Gamma}_{\mathrm{2D}}$ and the mean values for the position of the G-peak $\overline{\omega}_{\mathrm{G}}$ and of the 2D-peak $\overline{\omega}_{\mathrm{2D}}$ for all substrate materials investigated.}
\end{table}

{\textbf{Conclusion}}\\
In summary, we use spatially-resolved confocal Raman spectroscopy as a fast, non-invasive characterization tool to investigate the suitability of different material classes as substrates for high quality of graphene, while excluding external or transfer related degradations by protecting the graphene with hBN on top. From the Raman spectra, we extract strain, doping and the strain uniformity for the different substrate materials and present the data in colour-coded scatter plots, which allows to easily decide whether a potential substrate material is suitable for high-quality graphene devices. The presented study suggests that 2d materials such as hBN, WSe$_2$ and MoS$_2$ are well suited substrates for graphene, as they possess low values of $\Gamma_{\mathrm{2D}}$ and little overall doping. In contrast, oxidic substrates yield moderate values of $\Gamma_{\mathrm{2D}}$ and intermediate levels of charge carrier doping in graphene. These observations are in good agreement with typical results obtained from transport studies in these material systems\cite{Kre14}. In general, we observe a trend of decreasing strain uniformity with increasing substrate roughness. \\

\textbf{Acknowledgement}\\
We thank C. Neumann for helpful discussions and gratefully acknowledge support by the Helmholtz-Nanoelectronic-Facility (HNF), the DFG (SPP-1459), the ERC (GA-Nr. 280140), and the EU project Graphene Flagship (contract no. NECT-ICT-696656). Growth of hexagonal boron nitride crystals was supported by the Elemental Strategy Initiative conducted by the MEXT, Japan and JSPS KAKENHI Grant Numbers JP26248061,JP15K21722 and JP25106006.\\\\

\end{document}